\documentstyle[12pt]{article}
\newcommand{\be}{\begin{equation}}
\newcommand{\en}{\end{equation}}
\newcommand{\bea}{\begin{eqnarray}}
\newcommand{\ena}{\end{eqnarray}}

\newcommand{\hbo}{\hbox to 1 true cm {\hfill } }

\newcommand{\Tr}{\hbox{Tr}}

\def\dslash{\partial\kern-.5em\slash}
\def\kslash{k\kern-.5em\slash}

\begin{document}
\vglue 1truecm

\vbox{
UNIT\"U-THEP-8/1993 \hfill July 12, 1993
}

\vfil
\centerline{\bf \large Casimir effect of strongly interacting scalar
   fields$^1$}

\bigskip
\centerline{ K.\ Langfeld, F.\ Schm\"user, H.\ Reinhardt }
\medskip
\centerline{Institut f\"ur theoretische Physik, Universit\"at
   T\"ubingen}
\centerline{D--72076 T\"ubingen, Germany }
\bigskip

\vfil
\begin{abstract}

Non-trivial $\phi ^{4}$-theory is studied in a renormalisation group
invariant approach inside a box consisting of rectangular plates
and where the scalar modes satisfy periodic boundary conditions at
the plates. It is found that the Casimir energy exponentially approaches
the infinite volume limit, the decay rate  given by the scalar condensate.
It therefore  essentially differs
from the power law of a free theory. This might provide
experimental access to properties of the non-trivial vacuum.
At small interplate distances the system can no longer tolerate a
scalar condensate, and a first order phase transition to the
perturbative phase occurs. The dependence of the vacuum energy density
and the scalar condensate on the box dimensions are presented.

\end{abstract}

\vfil
\hrule width 5truecm
\vskip .2truecm
\begin{quote}
$^1$ Supported by DFG under contract Re $856/1 \, - \, 1$
\end{quote}
\eject
\section{ Introduction }
\label{sec:1}

\medskip
The Casimir effect~\cite{cas48,am83,plu86,bla88} in quantum field theory
is the change of the vacuum energy density due to constraints on
the quantum field induced by boundary conditions in space-time.
The contribution to the energy density by the quantum fluctuation of
the electromagnetic field was experimentally observed by
Sparnaay~\cite{spa58} in 1958, thus verifying its quantum nature.
Following this observation the Casimir effect was extensively studied,
the renormalisation procedure that must be used in order to extract
physical numbers out of divergent mode sums, being of particular
interest. This procedure is
most elegantly formulated in a path-integral approach~\cite{sym81},
and leads to a full understanding of the Casimir effect for
non-interacting quantum fields. Perturbative corrections
to the free Casimir arising from a weak interaction of the fluctuating
fields can also be obtained~\cite{to80}. It was shown that
the net effect of the
boundaries is to produce a topolocigal mass for the fluctuating
modes~\cite{fo79}. In the recent past there has been a
renaissance of the Casimir effect due to its broad span of applications,
which range from gravity models~\cite{gravity} to QCD bag
models~\cite{be76} to non-linear meson-theories describing baryons
as solitons~\cite{mou91}. A closely related subject is Quantum Field Theory
at finite temperature since it can be described in the path-integral
formalism by implementing periodic boundary conditions in Euclidean
time direction~\cite{do74}.
Despite these many different applications, it is possible to understand
the basic features of the Casimir effect by
investigating a scalar theory. It is also of general interest to study
$\phi ^{4}$-theory due to its important applications, e.g.\ in
the Weinberg-Salam model of weak interactions (see e.g.\ \cite{ch84})
and in solid state physics~\cite{stn72}. Many different
approaches~\cite{wet92,ste84} to strongly interacting $\phi ^{4}$-theory
were designed to understand its non-trivial vacuum structure.

Even though the Casimir effect of free quantum fields is well understood,
there is not yet an understanding of
the Casimir effect for strongly interacting fields.
This is simply due to the lack of knowledge of the true vacuum of an
interacting quantum theory. Recently, a non-perturbative path integral
approach to $\phi ^{4}$-theory has yielded
some insight into the vacuum structure of the strongly interacting
scalar theory~\cite{la92a,la92b}. In particular, it was found that
its perturbative phase is unstable (at zero temperature) because a
second phase with non-vanishing scalar condensate has lower vacuum
energy density~\cite{la92a}. In this phase, the connection between
the scalar condensate and the vacuum energy density, which is provided
by the scale anomaly, has been verified by an explicit
calculation~\cite{la92b}. The structure of this new
phase, describing strongly interacting scalar modes, was also investigated
at finite temperature~\cite{la92b}. It was found that at a critical
temperature the energy densities of the non-trivial and perturbative
phases are equal, and the non-trivial
phase undergoes a first order phase transition to the perturbative one.

Using these results it is possible to study the Casimir
effect of strongly interacting scalar fields. Since the non-trivial
phase provides an intrinsic energy scale (i.e.\ the magnitude of the
scalar condensate at zero temperature), one expects deviations of
the Casimir force from the free field law. This presumably provides
access to non-perturbative vacuum properties.

In this paper, we investigate the non-trivial phase of four dimensional
$\phi ^{4}$-theory in a rectangular box consisting of $p \; (<4)$
pairs of oppositely layered plates separated by a distance $a_{p}$,
with the scalar modes satisfying periodic boundary conditions at the plates.
We shall find at large interplate distances, that
the Casimir energy decays exponentially with increasing distance,
the decay rate given by the magnitude of the scalar
condensate. At small distances, the field theory no longer tolerates a
scalar condensate, and the perturbative phase is adopted.

The paper is organised as follows: in the second section we briefly
review the Casimir effect of a free theory and the recently
proposed non-perturbative approach~\cite{la92a,la92b}
to $\phi ^{4}$-theory. The renormalisation procedure is
discussed and renormalisation group invariance is shown.
In the subsequent section results are presented.
The Casimir energy as a function of large (compared with the scalar
condensate) interplate distance is obtained analytically, and
the deviations from the energy in a free field theory are discussed.
The phase-transition from the non-trivial vacuum to the perturbative phase
at small interplate distances is studied and the vacuum energy density and
the scalar condensate is calculated as function of the plate distances.
Discussions and concluding remarks are given in the final section.

\bigskip
\section{ The Casimir effect of scalar fields }
\label{sec:2}

\medskip
$\phi ^{4}$-theory is described by the Euclidean generating functional
for Green's functions, i.e.
\be
Z[j] \; = \; \int {\cal D} \phi \; \exp \{ - \int d^{4}x \;
\bigl( \frac{1}{2} \partial _{\mu } \phi \partial _{\mu } \phi \, + \,
\frac{ m^{2} }{2} \phi ^{2} \, + \, \frac{ \lambda }{24} \phi ^{4}
\, - \, j (x) \phi ^{2} (x) \, \bigr) \; \} \; ,
\label{eq:2.1}
\en
where $m$ denotes the bare mass of the scalar field and $\lambda $ the
bare coupling strength of the $\phi ^{4}$-interaction.
$j(x)$ is an external source for $\phi ^{2} (x)$ which is introduced
so that we can derive the
effective potential~\cite{co73,ja74} of the composite field $\phi ^{2}$
later on. It was observed in~\cite{la92b} that it is more convenient
to use the effective potential of $\phi ^{2}$ to study
the phase structure of the theory. In particular, its
minimum value is the vacuum energy density and thus provides access
to the Casimir effect, if it is calculated by imposing adequate
boundary conditions to the scalar modes.
For these initial investigations
we adopt the simplest geometry and consider
a rectangular box consisting of $p \; (<4)$ pairs of oppositely
layered plates separated by distances $a_{p}$. We expect
that the Casimir energy will not be sensitive to the detailed shape
of the finite volume as is known in a free theory~\cite{ba78}.
The integration over the field $\phi $ in (\ref{eq:2.1}), only extends
over configurations which satisfy periodic boundary conditions at the
plates. In the case of Dirichlet or
Neumann boundary conditions, surface counter terms must be added to
(\ref{eq:2.1}). In that case the results would sensitively depend on the
physical structure of the surface, and such effects are beyond the
scope of this paper.

The effective action is defined by a Legendre transformation of the
generating functional $Z[j]$, i.\ e.
\be
\Gamma [ \phi ^{2} _{c}] := - \ln Z[j] + \int d^{4}x \;
\phi ^{2}_{c}(x) j(x) \; , \hbo
\phi ^{2}_{c}(x) :=  \frac{ \delta \, \ln Z[j] }{ \delta j(x) } \; .
\label{eq:2.2}
\en
{}From here the effective potential $U(\phi ^{2}_{c})$ is obtained
by restricting $\phi _{c}$ to constant classical fields
($\Gamma [\phi _{c}^{2}=const.] = \int d^{4}x \; U(\phi _{c}^{2})$),
which are obtained
for a constant external source $j$. The minimum value of the
effective potential $U_{min}$ is the vacuum energy density and is
obtained from (\ref{eq:2.2}) at zero external source, i.e.\
\be
\frac{ d U }{ d \phi ^{2}_{c} } \vert _{\phi ^{2}_{c} = \phi ^{2}_{c \, 0 } }
\; = \; j \; = 0 \; .
\label{eq:2.3}
\en
The minimum classical configuration  $\phi ^{2}_{c \; 0 }$
represents the scalar condensate.

\bigskip
\subsection{Equivalence of effective action and sum of zero-point energies }
\label{sec:2.1}

\medskip
In this subsection we review the Casimir effect of a free scalar theory
$(\lambda =0)$ using Schwinger's proper-time regularisation.
We demonstrate that the minimum of the effective potential $U$
coincides with the mode sum usually considered when studying the Casimir
effect~\cite{cas48,am83,plu86,bla88}.
This equivalence was also obtained by using another regularisation
scheme~\cite{am83}, and previously observed with proper-time
regularisation in the context of chiral solitons~\cite{re89}.

The minimum of the effective potential of a free scalar theory is
\be
U_{min} \; = \;  \frac{1}{2 T V_{3}} \; \Tr \,
\ln ( - \partial ^{2} + m^{2} ) \; ,
\label{eq:2.4}
\en
where $T$ is the Euclidean time interval, and $V_{3}$ is the space volume.
The trace in (\ref{eq:2.4}), extending over all modes satisfying periodic
boundary conditions,  is a divergent object and needs
regularisation. For definiteness we use Schwinger's proper-time
regularisation, but note, however, that the specific choice
of the regularisation prescription has no
influence on the renormalised (finite) result (e.g., compare
\cite{ja74} and \cite{fo92}). In proper-time regularisation,
the vacuum energy density becomes
\be
U_{min} \; = \; - \frac{1}{2 T V_{3} } \Tr \int _{1/\Lambda ^{2} }^{\infty }
\frac{ ds }{ s } \; e^{ - s ( -\partial ^{2} + m^{2} ) } \; ,
\en
where $\Lambda $ is the ultraviolet cutoff.

The trace over the temporal degree of freedom can be easily performed, i.e.,
\be
U_{min} \; = \; - \frac{1}{2 V_{3}} \int \frac{ dk_{0} }{ 2 \pi }
\Tr_{V} \; \int _{1/ \Lambda ^{2} }^{\infty } \frac{ ds }{s} \;
e ^{ - s ( k_{0}^{2} + E ) } \; ,
\label{eq:2.5}
\en
The trace $\Tr_{V}$ extends over the spatial degrees of freedom and
$E$ is the energy obtained from the eigenvalue equation
\be
( - \nabla ^{2} + m^{2} ) \phi (x) = E^{2} \phi (x) \; ,
\nonumber
\en
where the eigenfunction $\phi $ satisfies periodic boundary conditions.
If the $k_{0}$-integra\-tion in (\ref{eq:2.5}) is performed,
a partial integration in the $s$-integral yields
\be
V_{3} U_{min} \; = \; \frac{1}{ 2 \sqrt{\pi } } \Tr_{V} \{ E \,
\Gamma ( \frac{1}{2}, \frac{ E^{2} }{ \Lambda ^{2} } ) \}
\; - \; \frac{ \Lambda }{ 2 \sqrt{\pi } }
\, \Tr_{V} \exp \{ - \frac{ E^{2} }{ \Lambda ^{2} } \} \; ,
\label{eq:2.6}
\en
where $\Gamma (\frac{1}{2},x)$ is the incomplete $\Gamma $-function.
The first term of (\ref{eq:2.6}) is precisely the mode sum
$\frac{1}{2} \Tr _{V} E $ in cutoff regularisation, with the particular
cutoff function $ \frac{1}{\sqrt{\pi }}
\Gamma ( \frac{1}{2}, \frac{ E^{2} }{ \Lambda ^{2} } ) $ provided by
Schwinger's proper-time regularisation. In the limit of large $\Lambda $,
the second term only contributes a constant to the action which is subtracted
by demanding that the Casimir energy approaches zero for large
interplate distances.

In order to illustrate the equivalence of the mode sum approach
and the approach provided by the effective potential,
we calculate the Casimir energy for a massless scalar particle
in a box consisting of $p$ pairs of rectangular plates and in
$d$ space-time dimensions.
In this case we have
\be
-\ln Z \; = \; - \frac{ L^{d-p} }{2}
\int \frac{ d^{d-p} k }{ (2\pi )^{d-p} }\;
\sum _{ \{ n_{i} \} = -\infty }^{\infty } \;
\int _{1/ \Lambda ^{2} } ^{\infty }
\frac{ ds }{s} \; \exp \{ -s [ k^{2} + \sum _{i=1}^{p} n_{i}^{2}
( \frac{ 2 \pi }{ a_{i} } )^{2} ] \} \; ,
\label{eq:2.7}
\en
where $a_{i}, i=1 \ldots p$ is the distance of the hyperplanes in $i$th
direction and $L \gg a_{i}$ is the length of the box of the unconstrained
modes. The integration over the continuous degrees of freedom
can be performed in a straightforward manner, i.e.,
\be
- \ln Z \; = \; - \frac{ L^{d-p} }{2} \frac{1}{ (2 \sqrt{\pi } )^{d-p} }
\sum _{ \{ n_{i} \} = -\infty }^{\infty } \;
\int _{1/ \Lambda ^{2} } ^{\infty }
\frac{ ds }{s ^{1+ \frac{d-p}{2} } } \; e^{ -s \sum n_{i}^{2}
( \frac{ 2 \pi }{ a_{i} })^{2}  } \; ,
\label{eq:2.7a}
\en
In order to extract the ultra-violet divergences, we apply
Poisson's formula
\be
\sum _{n=-\infty } ^{\infty } f(n) = \sum _{ \nu =-\infty } ^{\infty } c(\nu )
\; , \hbox to 2 true cm {\hfil with \hfil }
c(\nu ) = \int _{-\infty }^{\infty } dn \; f(n) \, e^{ i 2 \pi \nu  n } \; .
\label{eq:2.8}
\en
Rewriting
\be
\sum_{n=-\infty }^{\infty } e^{- s n^{2} (\frac{ 2 \pi }{ a })^{2} }
\; = \;
\frac{a}{ 2 \sqrt{\pi s } } \sum_{\nu=-\infty}^{\infty }
e^{- \nu^{2} \frac{ a ^{2} }{ 4 s }}
\label{eq:2.9}
\en
equation (\ref{eq:2.7a}) becomes
\be
- \ln Z \; = \; - \frac{ L^{d-p} }{2} \frac{1}{ (2 \sqrt{\pi } )^{d-p} }
\int _{1/ \Lambda ^{2} } ^{\infty }
\frac{ ds }{s ^{1+ \frac{d-p}{2} } } \;
\prod _{i=1}^{p} [ \frac{ a_{i} }{ 2 \sqrt{\pi s } }
\sum_{\nu _{i} =-\infty }^{\infty }
e^{- \nu _{i}^{2} \frac{ a _{i}^{2} }{ 4 s } } ] \; .
\label{eq:2.10}
\en
Note that the ultra-violet behaviour is dominated by the integrand
at small $s$ and the only divergences come from the term with
all $\nu _{i}$ are zero. The divergent term
\be
- \ln Z ^{div} \; = \; - \frac{ L^{d-p} }{2}
\frac{1}{ (2 \sqrt{\pi } )^{d-p} } \; \int _{1/ \Lambda ^{2} } ^{\infty }
\frac{ ds }{s ^{1+ \frac{d-p}{2} } } \;
\prod _{i=1}^{p} [ \frac{ a_{i} }{ 2 \sqrt{\pi s } }   ] \; ,
\label{eq:2.11}
\en
is proportional to the $d$-dimensional volume $V$ and a pure constant,
which can be absorbed by a redefinition of the action.
After the substitution $s \rightarrow 1/s$, the $s$-integral can be
performed in (\ref{eq:2.10}), yielding for the finite part in the
limit $\Lambda \rightarrow \infty $
\be
- \frac{1}{V} \ln Z \; = \; U_{min} \; = \; - \frac{ 1 }{ 2 }
\frac{ 1 }{ \pi ^{d/2} } \, \Gamma ( \frac{d}{2} )
\, Z(a_{1} \ldots a_{p}, d) \; ,
\label{eq:2.12}
\en
where
\be
Z( a_{1} \ldots a_{p}, d) = {\sum _{ \{ \nu _{i} \} = - \infty }^{ \infty } }
' \frac{1}{ (a_{1}^2 \nu_{1}^2 + \ldots + a_{p}^{2} \nu _{p}^{2}
)^{ \frac{d}{2} } } \; ,
\label{eq:2.13}
\en
is the Epstein Zeta-function (the prime indicates that the contribution with
all $\nu _{i}=0$ is excluded from the sum). For $p=1$ and four
space-time dimensions $(d=4)$ one obtains
the analytic result for the vacuum energy density and the
Casimir energy $E_{c}$, respectively, i.e.,
\be
U_{\min} \; = \; - \frac{ 1 }{ \pi ^{2} } \frac{1}{ a^{4} }
\Gamma ( 2 ) \zeta (4) \; , \hbo
E_{c} \; = \; V_{3} U_{min} \; = \; -\frac{ \pi ^{2} L^{2} }{90 a^{3} } \; ,
\label{eq:2.14}
\en
where $\zeta (s) = \sum_{1}^{\infty } n^{-s}$ is Riemann's
$\zeta $-function. This is precisely the result usually
obtained by evaluating the mode sum of zero point energies~\cite{am83}.

\subsection{Non-trivial $\phi ^{4}$-theory with boundary conditions }
\label{sec:2.2}

\medskip
In this subsection we describe the non-perturbative approach to
$\phi ^{4}$-theory provided by the modified loop expansion~\cite{la92a},
taking into account the constraints on the scalar field imposed by boundary
conditions. We demonstrate that the renormalisation procedure
is not affected by the presence of a rectangular box, implying
that renormalisation group invariance is preserved as it is in the
infinite volume limit ($a_{i} \rightarrow \infty $).

The modified loop expansion~\cite{la92a} is based on a linearisation
of the $\phi ^{4}$-interaction in the path-integral (\ref{eq:2.1})
by means of an auxiliary field $\chi (x)$
\bea
Z[j] \; = \; \int {\cal D} \phi \; {\cal D} \chi && \! \! \! \!
\exp \{ - \int d^{4}x \; [ \, \frac{1}{2} \partial_{\mu } \phi
\partial_{\mu } \phi \, +
\label{eq:2.15} \\
&& \frac{6}{\lambda } \chi^{2} (x)
\, + \, [\frac{ m^{2} }{2} - i \chi (x) ] \phi ^{2}(x)
\, - \, j(x) \phi ^{2}(x) ] \; \} \; .
\nonumber
\ena
This linearisation was first proposed in~\cite{co74}.
The integral over the fundamental field $\phi $ is then
easily performed, yielding
\bea
Z[j] &=& \int {\cal D} \chi \; \exp \{ - S[\chi,j] \} \; ,
\label{eq:3} \\
S[\chi,j] &=& \; \frac{ 6 }{ \lambda } \int d^{4}x \; \chi^{2}  \, + \,
\frac{1}{2} \Tr _{(R)} \ln \, {\cal D}^{-1}[\chi ,j] \; ,
\label{eq:2.16} \\
{\cal D}^{-1}[\chi ,j]_{xy} &=&
(- \partial ^{2} + m^{2} - 2i \chi (x) -2j(x)) \delta _{xy}  \; .
\label{eq:2.17}
\ena
The trace $\Tr _{(R)}$ extends over all eigenmodes of the operator
${\cal D}^{-1}[\chi ,j]$ which satisfy the periodic boundary conditions
and the subscript $(R)$ indicates that a regularisation prescription
is required. Note that the boundary conditions to the field $\phi $
do not give rise to any constraint for the auxiliary field $\chi $.
The approach of~\cite{la92a} is defined by a modified expansion with
respect to the field $\chi $ around its mean field value $\chi_{0}$
defined by
\be
\frac{ \delta S[ \chi , j] }{ \delta \chi (x) }
\vert _{\chi = \chi _{0} } \; = \; 0 \; .
\label{eq:2.18}
\en
The modified loop expansion of~\cite{la92a,la92b}
coincides with an $1/N$-expansion of
$O(N)$ symmetric $\phi ^{4}$-theory~\cite{ab76} for $N=1$, implying
that the convergence of the expansion is doubtful.
However, it was seen in zero dimensions that the effective potential of
this approximation rapidly converges to the exact one obtained numerically.
Recent results show that the same is true for four dimensional
$\phi ^{4}$-theory~\cite{lo93}. Reasonable results are obtained even at
mean-field level. At this level we obtain from (\ref{eq:2.15})
\bea
- \ln \, Z [j](a_{1} \ldots a_{p}) & = &
\label{eq:2.19} \\
& & \int d^{4}x \; \{ - \frac{3}{2 \lambda }
(M- m ^{2} + 2j)^{2} \} \; + \; \frac{1}{2} \Tr_{(R)} \, \ln (
- \partial ^{2} + M ) \; ,
\nonumber
\ena
where $M$ is related to the mean field value $\chi _{0}$ by
$\chi_{0}=i (M - m^{2} + 2j ) /2 $. The mean field equation for $\chi_{0}$
(\ref{eq:2.18}) can be recast into an equation for $M$, i.e.,
\be
\frac{ \delta \ln \, Z[j] }{ \delta M } \; = \; 0 \; .
\label{eq:2.20}
\en
For a constant external source $j$, this equation is satisfied for
constant $M$.

The only effect of the rectangular box is contained in the loop
contribution, which is, in Schwinger's proper-time regularisation
$(d=4)$
\bea
L &=& \frac{1}{2} \Tr_{(R)} \, \ln (- \partial ^{2} + M )
\label{eq:2.21} \\
& = &
- \frac{ L^{4-p} }{2} \int \frac{ d^{4-p} k }{ (2\pi )^{4-p} } \;
\sum _{\{n_{i}\}} \int _{1 / \Lambda ^{2} }^{\infty }  \frac{ds}{s} \;
\exp \{ -s [  \sum_{l=1}^{4-p} k_{l}^{2} + M + \sum_{i}
( \frac{ 2\pi }{ a_{i} } )^{2} n_{i}^{2} ] \} \; ,
\nonumber
\ena
with $i=1 \ldots p$ and $L$ the linear extension in the unconstrained
directions. The sum over the
unconstrained modes ($k$-integral) can be easily performed.
Applying Poisson's formula (\ref{eq:2.8}) to extract the
divergent terms as we did for the free theory (section \ref{sec:1}) we obtain
\be
L \; = \; - \frac{ V }{ 32 \pi ^{2} } M^{2} \, \Gamma (-2,
\frac{M}{ \Lambda ^{2} } ) \; - \; \frac{ V }{ 32 \pi ^{2} }
{\sum _{\{\nu_{i}\}} }'  \int _{ 1/ \Lambda ^{2} }^{\infty }
\frac{ ds }{s ^{3} } \;
e^{-s M } \; \exp (- \sum_{i} \frac{ a ^{2}_{i} }{ 4s } \nu_{i} ^{2} ) \; ,
\label{eq:2.22}
\en
where $\Gamma $ is the incomplete $\Gamma $-function, $V$ the
space-time volume and the prime
indicates that the contribution with all $\nu _{i}=0$ is excluded.
This implies that the second term of the right hand side of (\ref{eq:2.22})
is ultraviolet finite, so we can remove the regulator in this term
($\Lambda \rightarrow \infty $). Using the asymptotic expression of the
incomplete $\Gamma $-function, we find
\be
L \; = \; \frac{ V }{32 \pi ^{2} } \{ M \Lambda ^{2} + \frac{1}{2}
M^{2} ( \ln \frac{M}{\Lambda ^{2} } - \frac{3}{2} + \gamma ) \} \; - \;
\frac{ V }{32 \pi ^{2} }  \, F_{3}( M, a_{1} \ldots a_{p} ) \; ,
\label{eq:2.23}
\en
where $\gamma = 0.577...$ is Euler's constant and the function
$F_{3}$ is defined by
\be
F_{\epsilon }(M,a_{1} \ldots a_{p}) \; = \;
{\sum _{\{\nu_{i}\}} }'  \int _{0}^{\infty } \frac{ ds }{ s^{\epsilon } }
\; e^{-sM } \; e^{- \sum_{i}  \frac{ a^{2}_{i} }{4s} \nu_{i}^{2} }
\label{eq:2.24}
\en
with $\epsilon = 3$. The first term on the right hand side of (\ref{eq:2.23})
is precisely the effective potential in the infinite volume limit since the
function $F_{3}$ vanishes for $a_{i} \rightarrow \infty $.
The second term in (\ref{eq:2.23}) is thus the modification
of the effective potential
due to the presence of the plates. Note that this term is finite,
implying that the boundaries do not affect the renormalisation
procedure. This is the desired result.

Following the renormalisation scheme given in~\cite{la92b}, we absorb the
divergences in the bare parameters $\lambda, m, j $ by setting
\bea
\frac{6}{ \lambda } \; + \; \frac{1}{16 \pi ^{2} } \, (\ln
\frac{\Lambda ^{2}}{\mu ^{2}}  - \gamma +1 ) &=& \frac{6}{ \lambda _{R} }
\label{eq:2.25} \\
\frac{6}{\lambda } j \, - \, \frac{3 m^{2} }{\lambda } \; - \;
\frac{1}{32 \pi ^{2} } \Lambda ^{2} &=& \frac{6}{ \lambda _{R} } j_{R}
\, - \, \frac{ 3 m_{R}^{2} }{ \lambda _{R} }
\label{eq:2.26} \\
j \; - \; m^{2} &=& 0 \; ,
\label{eq:2.27}
\ena
where $\mu $ is an arbitrary renormalisation point and a subscript $R$
refers to the renormalised quantities. Later
we will check that physical quantities do not depend on $\mu $.
In the following, we consider the massless case $m_{R}=0$.
The coupling strength renormalisation in (\ref{eq:2.25})
was earlier used by Coleman et al.~\cite{co74} and, as pointed
out by Stevenson, it implies that the bare coupling
becomes (infinitesimally) negative, if the regulator $\Lambda $
is taken to infinity~\cite{ste84}. It was shown that this
behaviour of the bare coupling strength is hidden in the standard
perturbation theory~\cite{la92a}.
In fact, we have from (\ref{eq:2.25})
\be
\lambda \; = \; \frac{ \lambda _{R} }{ 1 - \beta _{0} \lambda _{R}
(\ln \frac{ \Lambda ^{2} }{ \mu ^{2} } - \gamma +1 ) } \; , \hbo
\beta _{0} \; = \; \frac{ 1 }{ 96 \pi ^{2} }
\label{eq:2.27a}
\en
implying $\lambda \rightarrow 0^{-}$ for $\Lambda \rightarrow \infty $,
whereas in contrast an expansion of (\ref{eq:2.27a}) with respect to the
renormalised coupling strength, i.e.
\be
\lambda \; = \; \lambda _{R}(\mu ) \, [ 1 \; + \;
\beta _{0} \, \lambda _{R}
( \ln \frac{ \Lambda ^{2} }{ \mu ^{2} } - \gamma +1 ) \; + \;
O( \lambda _{R} ^{2} ) ]
\en
suggests that $\lambda \rightarrow + \infty $, if $\Lambda
\rightarrow \infty $.

Inserting (\ref{eq:2.25}-\ref{eq:2.27}) and $L$ from (\ref{eq:2.23})
in (\ref{eq:2.19}) one obtains
\bea
-\frac{1}{ V } \ln \, Z[j](a_{1} \ldots a_{p} ) &=&
- \frac{3}{2 \lambda _{R} } M^{2} \, - \,
\frac{6}{\lambda _{R} } M j_{R} \, + \, \frac{ \alpha }{ 2 }
M^{2} ( \ln \frac{M}{\mu ^{2} } - \frac{1}{2} )
\label{eq:2.28} \\
&-&   \alpha \, F_{3}( M, a_{1} \ldots a_{p})
\nonumber
\ena
where $\alpha = 1/ 32 \pi ^{2}$, and $M$ is defined by the mean field
equation (\ref{eq:2.20}), i.e.
\be
- \frac{3}{\lambda _{R} } M \, - \, \frac{6}{\lambda _{R} } j_{R}
\, + \, \alpha M ( \ln \frac{M}{\mu ^{2} } - \frac{1}{2} ) \, + \,
\alpha \, F_{2}( M,a_{1} \ldots a_{p} ) \; = \; 0 \; .
\label{eq:2.29}
\en
It is now straightforward to perform the Legendre transformation
(\ref{eq:2.2}). The final result for the effective potential is
\be
U(\phi _{c}^{2}) \; = \; \frac{\alpha }{2} M^{2} (
\ln \frac{M}{\mu ^{2} } - \frac{1}{2} ) \; - \;
\frac{3}{2 \lambda _{R} } M ^{2} \; - \; \alpha \,
F_{3}( M, a_{1} \ldots a_{p} ) \; ,
\label{eq:2.30}
\en
where
\be
\phi ^{2}_{c} \; = \; \frac{1}{ V }
\frac{ \delta \ln Z[j_{R}] }{\delta j_{R} } \; = \;
\frac{6}{\lambda _{R}} M \; .
\label{eq:2.31}
\en
The effective potential is renormalisation group invariant, since
a change in the renormalisation point $\mu $ can be absorbed
by a change of the renormalised coupling strength~\cite{la92a,la92b}.
Note that due to field renormalisation (\ref{eq:2.27}),
$M= \lambda _{R} \phi ^{2}_{c}/6$ is renormalisation group invariant
rather than $\phi ^{2}_{c}$. Thus $M$ is a physical quantity and is
referred to as scalar condensate. In the infinite volume limit
$(a_{i} \rightarrow \infty )$ the effective potential has a global
minimum for $M=M_{0} \not= 0$, implying that the ground state
has a non-vanishing scalar condensate~\cite{la92b}. Furthermore,
the minimum value of the effective potential (vacuum energy density)
is related to the scalar condensate by~\cite{la92b}
\be
U_{0} \; = \; - \frac{\alpha }{144} \lambda^{2}_{R} (\phi_{c}^{2})^{2}
\; \rightarrow \; - \frac{1}{4}
\frac{ \beta (\lambda _{R}) }{24} \langle :\phi ^{4}: \rangle \; ,
\label{eq:2.32}
\en
which yields the correct scale anomaly at this level of approximation.

In order to make renormalisation group invariance obvious, we
remove the renormalisation point dependence in (\ref{eq:2.30}) by
subtracting $U_{0}$ from the effective potential $U(M)$ (\ref{eq:2.30}).
Both the renormalisation point $\mu $ and the renormalised
coupling $\lambda _{R}$ drop out, and we obtain
\be
U(M) \; = \; \frac{\alpha }{2} M^{2} ( \ln \frac{M}{M_{0}} - \frac{1}{2} )
\; - \; \alpha F_{3}(M, a_{1} \ldots a_{p}) \; .
\label{eq:2.33}
\en
The effective potential $U$ as a function of the scalar condensate $M$
for one pair of plates ($p=1$) is shown in figure 1.
In this case, the results  are equivalent to that of a finite temperature
field theory if the inverse distance $1/a$ between the plates
is identified with the temperature (in units of Boltzmann's
constant)~\cite{do74}.
Further results of finite temperature $\phi ^{4}$-theory
are given in~\cite{la92b}.
For a large interplate distance $a$ (zero temperature), the continuum
effective potential is obtained, and the effective potential has
a minimum at a nonvanishing value of the scalar condensate. At finite $a$,
a second minimum at zero condensate $M$ develops, which is referred to as
the perturbative phase. At large $a$, this trivial phase is unstable, because
the non-perturbative minimum has lower vacuum energy density.
Decreasing $a$ (increasing temperature), lowers the difference in the
energy density between the perturbative and non-perturbative phases.
At a critical distance $a_{c}$ the non-trivial phase becomes degenerate
with the perturbative one (at $M=0$). If $a$ is decreased further,
the non-trivial phase becomes meta-stable and a first order
phase transition to the trivial phase at $M=0$ can
occur, either by quantum or statistical fluctuations.

\bigskip
\section{ Results }

\medskip
In a free massless field theory (with $p=1$) there is no intrinsic energy
scale in competition with that of the interplate distance.
This implies that the vacuum energy density $U_{0}$ scales
as $1/a^{4}$ with the interplate distance from dimensional
arguments. This scaling law was experimentally observed by
Sparnaay in the case of QED~\cite{spa58}. In the case of
non-trivial $\phi ^{4}$-theory an intrinsic energy scale is
provided by the scalar condensate. Thus one expects deviations
from the $1/a^{4}$ scaling law of the free theory. Such deviations
might provide experimental access to properties of the non-trivial
phase.

\subsection{ Near the infinite volume limit }
\label{sec:3.1}

\medskip

The scalar condensate $M_{v}$ at the minimum of the effective potential
is given by the gap equation
\be
\frac{ dU }{ dM } \vert_{ M=M_{v}} \; = \;
M_{v} \, \ln \frac{ M_{v} }{ M_{0} } \; + \; F_{2}( M_{v}, a_{1}
\ldots a_{p} ) \; = \; 0 \; .
\label{eq:2.34}
\en
The vacuum energy density $U_{v}$ is obtained by inserting $M_{v}$
back into (\ref{eq:2.33}).
For large interplate separations $a_{i}^{2} \gg 1/M_{v}$ the function
$F_{\epsilon }( M_{v}, a_{1} \ldots a_{p})$ can be analytically
estimated by noting that only terms with a single $\nu _{i} \not= 0$
and all others $\nu _{j \not= i} =0$ contribute to the sum
(\ref{eq:2.24}), i.e.
\be
F_{\epsilon }(M,a_{1} \ldots a_{p}) \; \approx \;
\sum _{i}   \int _{0}^{\infty } \frac{ ds }{ s^{\epsilon } }
\; e^{-sM } \; e^{- \frac{ a^{2}_{i} }{4s} } \; .
\label{eq:2.35}
\en
A simple rescaling yields
\be
F_{\epsilon }(M,a_{1} \ldots a_{p}) \; = \; \sum _{i}
(\frac{4}{ a_{i} ^{2} })^{\epsilon  -1} \, f_{\epsilon } ( \frac{
M_{v} a_{i}^{2} }{4} ) \; , \hbo
f_{\epsilon }(x) \; = \;  \int _{0}^{\infty } \frac{ ds }{ s^{\epsilon } }
\; e^{-s x } \; e^{- \frac{ 1 }{s} } \; .
\label{eq:2.36}
\en
After some technical manipulations, the functions $f_{\epsilon }(x)$
can be related to the modified Bessel functions of the second kind, i.e.,
\be
f_{\epsilon }(x) \; = \; 2 \, x^{ \frac{ \epsilon -1 }{2} } \,
K_{\epsilon -1 }( 2 \sqrt{x} ) \; \approx \;
\sqrt{ \pi } \, x^{ \frac{ 2 \epsilon -3 }{4} } \, e^{ -2 \sqrt{x} } \; ,
\label{eq:2.37}
\en
where the last approximate expression is just the asymptotic form of the
Bessel function for $x \rightarrow \infty $. Thus we have near the
infinite volume limit
\be
F_{2} \; \approx \; M^{1/4} \sqrt{\pi} \sum _{i} (
\frac{ a_{i}^{2} }{4} )^{-3/4} \, e^{-\sqrt{M} a_{i} } \; , \hbo
F_{3} \; \approx \; M^{3/4} \sqrt{\pi} \sum _{i} (
\frac{ a_{i}^{2} }{4} )^{-5/4} \, e^{-\sqrt{M} a_{i} } \; .
\label{eq:2.38}
\en
Solving (\ref{eq:2.34}) for the scalar condensate $M_{v}$ in perturbation
theory around $M_{0}$ we obtain the change in the condensate due to
the presence of the plates, i.e.
\be
M_{v} \; = \; M_{0} \, [ 1 \; - \; \sqrt{\pi } \, \sum _{i=1}^{p}
(\frac{ M_{0} a_{i}^{2} }{4} ) ^{-3/4} \, e^{- \sqrt{ M_{0} } a_{i} }
\; + \; \ldots \; ] \; .
\label{eq:2.39}
\en
There are two contributions to the variation of the vacuum energy density
\break
$U_{v}( M_{v}(a_{i}), a_{i} )$
(\ref{eq:2.33}) with the interplate distances, one from a change of
the scalar condensate and one from the change of the effective
potential $U(M)$ via the function $F_{3}$. Equation (\ref{eq:2.34})
implies that a variation of the condensate does not change $U_{v}$
in first order, and thus the leading contribution is from a change
of $F_{3}$. Using the asymptotic form
(\ref{eq:2.38}) for $F_{3}$ one obtains
\be
\frac{1}{\alpha } \Delta U_{v} \; = \; \frac{1}{\alpha }
(U_{v} - U_{\infty }) \; \approx \;
- \sqrt{\pi } \, M_{0}^{2} \, \sum_{i=1}^{p}
( \frac{ M_{0} a_{i}^{2} }{4} )^{-5/4} \, e^{- \sqrt{M_{0}} a_{i} } \; ,
\label{eq:2.40}
\en
where $U_{\infty }$ is the vacuum energy density in the infinite
volume limit.
This is the desired result: equation (\ref{eq:2.40}) gives the change
of the vacuum energy density due to boundary conditions.
In free field theory (and $p=1$) it decays by the power law $\sim 1/a^{4}$
(see (\ref{eq:2.14})). In contrast, the energy density (\ref{eq:2.40}) of
strongly interacting scalar modes decays exponentially (with a power law
correction), the slope given by the magnitude of the scalar
condensate $M_{0}$.
This implies that at least in principle one can decide by observing
the dependence of the Casimir force on the interplate distance,
whether the theory is in a free or in a non-perturbative phase.
In the latter case, it is also possible to extract ground state
properties, e.g.\ the scalar condensate. Since in QED the
$1/a^{4}$-power law was experimentally verified~\cite{spa58},
the QED ground state is trivial and e.g., has no photon condensate,
an expected result since photon self-interactions are absent.

\bigskip
\subsection{ At the phase transition at small interplate distances }
\label{sec:3.2}

\medskip
As was seen in section \ref{sec:2.2} for one pair of plates $(p=1)$,
the system undergoes a first order phase transition from the non-trivial
vacuum to the perturbative vacuum, if the interplate distance becomes
small enough. Numerical investigations of the effective potential
(\ref{eq:2.33}) at various distances $a_{i}$ show that the same
effect holds for $p>1$: if the box is small enough, a first order
phase transition to the perturbative vaccum occurs.

Equating the energy density $U_{0}$  of the perturbative state
at $M=0$  to that  of the non-trivial phase (with non-zero
condensate) at $M=M_{v}$ we obtain
\be
\frac{ M_{v}^{2} }{ 2 } ( \ln \frac{ M_{v} }{ M_{0} } - \frac{1}{2} )
\, - \, F_{3}( M_{v}, a_{1} \ldots a_{p}) \, + \,
F_{3}( 0, a_{1} \ldots a_{p}) \; = \; 0 \; ,
\label{eq:2.41}
\en
where the dependence of the scalar condensate $M_{v}(a_{1} \ldots a_{p})$
on the interplate distances $a_{i}$ is implicitly given by (\ref{eq:2.34}).
The set of equations (\ref{eq:2.34}, \ref{eq:2.41}) defines a
hypersurface in the space spanned by the distances $a_{i}$, which
separates the non-trivial phase from the perturbative one.
Note that this transition line is given in terms of renormalisation
group invariant (and therefore physical) quantities.

For {\it one pair \/} of plates ($p=1$), the formulation is equivalent
to the finite-temperature $\phi ^{4}$-theory (identifying $1/a$
with temperature), and the phase transition at small distance $a$
is of same structure as that in the finite-temperature theory at
high temperature.
Due to this correspondence, the numerical value for the critical
distance $a_{(c)}$ can be taken from~\cite{la92b}
\be
M_{0} \, a^{2}_{(c)} \; = \; 10.29134 \ldots \; .
\label{eq:2.42}
\en
The ratio of the scalar condensate $M_{v}$ at the transition point
and the continuum (zero temperature) condensate $M_{0}$ is
\be
M_{v}(a_{(c)}) \, / \, M_{0} \; = \; 0.9041 \ldots \; .
\label{eq:2.43}
\en
Due to the first order nature of the phase transition,
the scalar condensate has a discontinuity at the transition point $a_{(c)}$
and is zero for smaller distances.

For {\it two pairs \/} of plates the transition line between the
two phases was obtained
by solving (\ref{eq:2.34}, \ref{eq:2.41}) numerically.
The result is presented in figure 2. Numerical investigations
(cf.\ figure 4) suggest that the transition line
is approximately given by the equation
\be
\frac{1}{ a_{c \, 1 }^{2} } \, + \, \frac{1}{a_{c \, 2 }^{2} } \; = \;
\frac{M_{0}}{9} \; .
\label{eq:2.44}
\en
For $p=3$ we have numerically checked, that the first order phase
transition occurs, if the rectangular box is sufficiently small.

\bigskip
\subsection{ Boundary dependence of energy density and scalar condensate }
\label{sec:3.3}

\medskip
For given vacuum energy density $U_{v}$, equation (\ref{eq:2.33}) i.e.,
\be
\frac{ M_{v}^{2} }{ 2 } ( \ln \frac{ M_{v} }{ M_{0} } - \frac{1}{2} )
\, - \, F_{3}( M_{v}, a_{1} \ldots a_{p}) \; = \;
U_{v}
\label{eq:2.45}
\en
with $M_{v}$ defined by (\ref{eq:2.34})
yields the hypersurface of constant energy density in the space
spanned by $\{a_{i}, i=1 \ldots p \}$.
Comparing (\ref{eq:2.45}) with (\ref{eq:2.41}) it is easily seen that
the phase separating surface is not a surface of constant energy
density, implying that there are intersections between the two surfaces.
We expect
that the hypersurface of constant energy density is continuous
at the intersection, but not differentiable due to the
first order phase transition.

Figure 3 shows the vacuum energy density for one pair
of plates ($p=1$) as a function of $16/a^{2}$, when $a$ is the interplate
distance (or equivalently the inverse temperature). For large
values of $16/a^{2}$ (small $a$), the perturbative phase is realised,
and the $1/a^{4}$-scaling law is observed. For small values of $16/a^{2}$
(large $a$) the scalar theory is in the non-trivial phase, and
the energy density exponentially approaches the continuum value
(see (\ref{eq:2.40})) given by the scale anomaly (\ref{eq:2.32}).

For two pairs of plates $(p=2)$, figure 4 shows lines of constant
vacuum energy density in the $16/a_{1}^{2} - 16/a_{2}^{2}$
plane. Also shown is the phase transition line (dashed curve).
The lines of constant energy density are continuous, but have a cusp
at the first order phase transition point.

We have also studied the hypersurfaces of constant
{\it scalar condensate \/ } in $a_{i}$-space. For $p=1$,
this is equivalent investigating the temperature dependence of the
scalar condensate, and thus the results are given in~\cite{la92b}.
For $p=2$, the lines of constant condensate (in units of continuum
condensate $M_{0}$) are presented in figure 5. A line
of constant condensate is discontinuous at the phase
transition line, and is zero in the perturbative phase.
This behaviour is again due to the first order phase transition.

\bigskip
\section{ Discussion and concluding remarks }
\label{sec:3}

\medskip
We have shown for $\phi^{4}$-theory constrained by a rectangular box
that the non-trivial ground state undergoes a first order phase
transition to the perturbative vacuum, if the extension in at least
one space-time direction becomes small enough. For
large boxes the finite size corrections to the infinite volume limit
are exponentially small. They are negligible, if the interplate
distances are large compared with intrinsic scale provided by the
(continuum) scalar condensate (i.e.\ $a_{i} \sqrt{M_{0}} \gg 1$).
On the other hand finite size effects become important for
$a_{i} \sqrt{M_{0}}
\approx 1$ and induces a phase transition to a non-trivial vacuum.

We believe that these properties are a common feature of a wide class
of quantum field theories. Indeed, an analogous situation is
observed in lattice gauge theories. Theoretical investigations show,
that for high temperatures pure SU(N) lattice gauge theory
has a phase transition from a non-trivial (confining) ground state
to a perturbative phase~\cite{pol78}. Numerical simulations of
the SU(N) theory use a lattice with size
$n_{t} n^{3}, \, n \gg n_{t} $, which corresponds to a system with volume
$n^{3}$ and inverse temperature $n_{t}$. Such a system shows two phases,
a non-trivial phase for $\beta < \beta_{c}(n_{t})$, and a deconfined
phase for $\beta > \beta_{c}(n_{t})$~\cite{ber88} ($\beta = 2N /g^{2}$
with $g$ the SU(N) coupling strength).
This compares well with our considerations as follows. The intrinsic scale
of lattice gauge theories is provided by the string tension
$\chi $~\cite{cr80} (or equivalently by the gauge field
condensate~\cite{gi81} as in the continuum Yang-Mills theory). Our
investigations suggest that a finite size phase transition occurs if
\be
n_{t} n^{3} \; a^{4} \chi ^{2} \; \le \; 1 \; ,
\label{eq:2.46}
\en
where $a$ is the lattice spacing.
The string tension in units of the lattice spacing $a^{4}\chi^{2}$,
strongly depends on the inverse coupling strength $\beta $
dictated by the renormalisation group. Numerical simulations~\cite{cr80}
show that for fixed $n_{t},n$, $a^{4} \chi ^{2}$
decreases with increasing $\beta $, implying that (\ref{eq:2.46}) is
satisfied for $\beta \approx \beta _{c}$, the coupling strength at which
the phase transition occurs.

\medskip
In conclusion, we have studied $\phi ^{4}$-theory
in a renormalisation group invariant approach inside a
rectangular box consisting of $p$ pairs of plates, at which the
scalar modes satisfy periodic boundary conditions.
We have further investigated the ground state properties of the non-trivial
phase affected by the geometrical constraints.
The dependence of the vacuum energy density and the scalar condensate
on the interplate distances was studied in some detail.
In the non-trivial phase the vacuum energy density exponentially
approaches the infinite volume limit, the decay rate given by the
magnitude of the scalar condensate.
This behaviour of the energy density essentially differs form that
of a free theory, where it scales according a $1/a^{4}$-power law.
This implies, that at least in principle, one can determine which
phase the system has adopted by measuring the Casimir force.
At small interplate distances, the system undergoes a first order
phase transition to the perturbative phase. This phase transition is
of the same nature as the transition at high temperature.

\bigskip
\leftline{\bf Acknowledgements: }

We want to thank R.\ F.\ Langbein for a careful reading of this
manuscript and useful remarks.
\medskip

\vfill \eject
\centerline{ \Large Figure captions }
\vspace{2cm}
{\bf Figure 1: } The effective potential as function of the scalar condensate
   at various interplate distances.

\vspace{1cm}
{\bf Figure 2: } The transition line separating the non-trivial phase and
   the perturbative phase, two pairs pf plates $(p=2)$ with distances
   $a_{1}$ and $a_{2}$, respectively.

\vspace{1cm}
{\bf Figure 3: } The vacuum energy density for one pair of plates
$(p=1)$ as a function of the
interplate distance (inverse temperature) in units of $1/ \sqrt{ M_{0} }$.

\vspace{1cm}
{\bf Figure 4: } The lines of constant vacuum energy density
$ \frac{ U_{v} }{ \alpha M_{0}^{2} }$ for two pairs of plates;
$a^{2}_{i}$ in units of the inverse scalar condensate $1/M_{0}$.

\vspace{1cm}
{\bf Figure 5: } The lines of constant scalar condensate $M/M_{0}$
for two pairs of plates.


\begin{thebibliography}{sch90}
\bibitem{cas48}{ H.\ B.\ G.\ Casimir, Proc.K.Ned.Akad.Wet. 51(1948)793. \\
   M.\ Fierz, Helv.Phys. Acta 33(1960)855. }
\bibitem{am83}{ J.\ Ambjorn, S.\ Wolfram, Ann.\ of Phys. 147(1983)1. }
\bibitem{plu86}{ G.\ Plunien, B.\ M\"uller, W.\ Greiner,
   Phys.Rep. 134(1986)88. }
\bibitem{bla88}{ S.\ K.\ Blau, M.\ Visser, Nucl.Phys. B310(1988)163. }
\bibitem{spa58}{ M.\ J.\ Sparnaay, Physica 24(1958)751. }
\bibitem{sym81}{ K.\ Symanzik, Nucl.Phys. B190(1981)1. }
\bibitem{to80}{ D.\ J.\ Toms, Phys.Rev. D21(1980)2805. }
\bibitem{fo79}{ L.\ H.\ Ford, T.\ Yoshimura, Phys.Lett. A70(1979)89. \\
   D.\ J.\ Toms, Phys.Rev. D21(1980)928. }
\bibitem{gravity}{ Osamu Abe, Prog.Theor.Phys. 72(1984)1225. \\
   Y.\ Igarashi, T.\ Nonoyama, Phys.Lett. B161(1985)103. \\
   Chang-Jun Ahn, Won-Tae Kim, Young-Jai Park, Kee Yong Kim,
   Yongduk Kim, Mod.Phys.Lett. A7(1992)2263. }
\bibitem{be76}{ C.\ M.\ Bender, P.\ Hays, Phys.Rev. D14(1976)2622. \\
   K.\ A.\ Milton, Phys.Rev D22(1980)1441, D27(1983)439,
   Ann. of Phys. 150(1983)432. }
\bibitem{mou91}{ H.\ Reinhardt, R.\ W\"unsch, Phys.Lett. B215(1988)577. \\
   R.\ Alkofer, H.\ Reinhardt, H.\ Weigel, U.\ Z\"uckert, \\
   Phys.Rev.Lett. 69(1992)1874. \\
   H.\ Weigel, R.\ Alkofer, H.\ Reinhardt, Nucl.Phys. B387(1992)638. \\
   B.\ Moussallam, D.\ Kalafatis, Phys.Lett. B272(1991)196. \\
   G.\ Holzwarth, Phys.Lett. B291(1992)218. }
\bibitem{do74}{ L.\ Dolan, R.\ Jackiw, Phys.Rev. D(1974)9. }
\bibitem{ch84}{ Ta-Pei Cheng, Ling-Fong Li, 'Gauge Theory of Elementary
  Particle \\ Physics', Oxford University Press, New York 1984. }
\bibitem{stn72}{ H.\ E.\ Stanley, Introduction to Phase Transitions and
   Critical Phenomena, Oxford University Press, London, 1972. \\
   P.\ G.\ de Gennes, Scaling Concepts in Polymer Physics,
   Cornell University Press, 1979. }
\bibitem{wet92}{ N.\ Tetradis, C.\ Wetterich, Nucl.Phys. B398(1993)659. \\
   M.\ B.\ Einhorn, D.\ R.\ T.\ Jones, Nucl.Phys. B398(1993)611. \\
   C.\  Wetterich, Nucl.Phys., B352(1991)529. \\
   A.\ Ringwald, C.\ Wetterich, Nucl.Phys. B334(1990)506. }
\bibitem{ste84}{ M.\ Stevenson, Z.\ Phys.\ C24(1984)87. \\
   M.\ Stevenson, Phys.Rev. D32(1985)1389. \\
   M.\ Stevenson, R.\ Tarrach, Phys. Lett. B176 (1986) 436.}
\bibitem{la92a}{K.\ Langfeld, L.\ v.\ Smekal, H.\ Reinhardt,
   Phys.Lett. B308(1993)279. }
\bibitem{la92b}{K.\ Langfeld, L.\ v.\ Smekal, H.\ Reinhardt,
   'Non-trivial phase structure of $\phi ^{4}$-theory at finite
   temperature', Phys.Lett. B in press. }
\bibitem{co73}{ S.\ Coleman, E.\ Weinberg, Phys.Rev. D7(1973)1888. \\
   S.\ Coleman, D.\ J.\ Gross, Phys.Rev.Lett. 31(1973)851. }
\bibitem{ja74}{ R.\ Jackiw, Phys.Rev. D9(1974)1686. }
\bibitem{ba78}{ R.\ Balian, B.\ Duplantier, Ann.Phys. 112(1978)165. \\
   T.\ H.\ Boyer, Phys. Rev. 174(1968)1764. \\
   B.\ Davies, J.Math.Phys. 13(1972)1324. \\
   W.\ Lukosz, Physica 56(1971)109. \\
   G.\ Cognola, L.\ Vanzo, S.\ Zerbini, J.Math.Phys. 33(1992)222. }
\bibitem{re89}{ H.\ Reinhardt, Nucl.Phys. A503(1989)825. }
\bibitem{fo92}{ C.\ Ford, D.\ R.\ T.\ Jones, Phys.Lett. B274(1992)409. }
\bibitem{co74}{ S.\ Coleman, R.\ Jackiw, H.\ D.\ Politzer,
   Phys.Rev. D10(1974)2491. }
\bibitem{ab76}{L.\ F.\ Abbot, J.\ S.\ Kang, H.\ J.\ Schnitzer,
   Phys.Rev. D13(1976)2212.}
\bibitem{lo93}{ L.\ v.\ Smekal, K.\ Langfeld, H.\ Reinhardt,
   'Scaling improved $1/N$-expansion of non-trivial $\phi ^{4}$-theory',
   in preparation. }
\bibitem{pol78}{ A.\ M.\ Polyakov, Phys.Lett. B72(1978)477. \\
   L.\ Susskind, Phys.Rev D20(1979)2610. }
\bibitem{ber88}{ B.\ A.\ Berg, A.\ Billoire, R.\ Salvador,
   Phys.Rev. D37(1988)3774. \\
   B.\ Petersson, Nucl.Phys. B(Proc.Suppl.)30(1993)66. \\
   D.\ Toussaint, Nucl.Phys. B(Proc.Suppl.)26(1992)3. \\
   B.\ Petersson, Nucl.Phys. A525(1991)237c. }
\bibitem{cr80}{ M.\ Creutz, Phys.Rev. D21(1980)2308. \\
  M.\ Creutz, Phys.Rev.Lett. 45(1980)313. \\
  G.\ M\"unster, Phys.Lett. B95(1980)59. }
\bibitem{gi81}{ A.Di Giacomo, G.C.\ Rossi, Phys.Lett. B100(1981)481. }
\end{thebibliography}
\end{document}